\documentclass[prd,preprint,superscriptaddress,preprintnumbers,eqsecnum,showpacs,nofootinbib,nobibnotes]{revtex4}
\usepackage{amsfonts,bm,mathbbol}
\usepackage{graphicx}
\usepackage{pstricks}

\newcommand{\be}{\begin{equation}}
\newcommand{\bea}{\begin{eqnarray}}
\newcommand{\ee}{\end{equation}}
\newcommand{\eea}{\end{eqnarray}}

\def\1eq#1{Eq.~(\ref{#1})}
\def\2eqs#1#2{Eqs.~(\ref{#1}) and~(\ref{#2})}
\def\3eqs#1#2#3{Eqs.~(\ref{#1}), (\ref{#2}) and~(\ref{#3})}
\def\4eqs#1#2#3#4{Eqs.~(\ref{#1}), (\ref{#2}), (\ref{#3}) and~(\ref{#4})}

\def\chic#1{{\scriptscriptstyle #1}}
\def\ie{{\it i.e.}, }

\begin{document}

\title{Ghost propagator and ghost-gluon vertex \\ from Schwinger-Dyson equations}
%
\author{A.~C. Aguilar}
\affiliation{University of Campinas - UNICAMP, Institute of Physics ``Gleb Wataghin'',
13083-859 Campinas, SP, Brazil}
\author{D.~Ib\'a\~nez}
%
%
\author{J. Papavassiliou}
\affiliation{ Department of Theoretical Physics and IFIC,
University of Valencia and CSIC,\\
E-46100, Valencia, Spain}

\begin{abstract}

We study an approximate version of 
the  Schwinger-Dyson equation that controls the 
nonperturbative behavior of the ghost-gluon vertex, 
in the Landau gauge. In particular, 
we focus on the form factor that 
enters in the 
dynamical equation for the ghost dressing function, in the same gauge, 
and derive its integral equation, in the 
``one-loop dressed'' approximation. 
We consider two 
special kinematic configurations, which simplify the 
momentum dependence of the unknown quantity; 
in particular, 
we study the soft gluon case, and the well-known Taylor limit.   
When coupled with the  
Schwinger-Dyson equation of the  ghost dressing function,
the contribution of this  
form factor provides considerable support to the relevant integral kernel.
As a consequence, the solution of this coupled system of integral equations 
furnishes a ghost dressing function that reproduces 
the standard lattice results rather accurately, 
without the need to artificially increase the 
value of the gauge coupling.

\end{abstract}


\maketitle

\section{Introduction}
One of the few nonperturbative frameworks available 
for the study of the  infrared sector of QCD in the continuum are the 
Schwinger-Dyson equations (SDEs), which govern the dynamics 
of the basic Green's functions of the theory~\cite{Roberts:1994dr,Alkofer:2000wg,Fischer:2006ub,Binosi:2009qm}. 
Despite the well-known limitations intrinsic to this formalism, a variety of 
theoretical and technical advances have provided new valuable insights 
on some of the most fundamental nonperturbative phenomena of QCD,
such as quark confinement, dynamical mass generation, 
and chiral symmetry breaking~\cite{Szczepaniak:2001rg,Szczepaniak:2003ve,Fischer:2006ub,Aguilar:2008xm,Aguilar:2010cn}.  
Particularly important in this ongoing effort is the 
systematic confrontation of the  SDE predictions with 
the results of large-volume lattice simulations~\cite{Cucchieri:2007md,Bowman:2007du,Bogolubsky:2007ud,Bogolubsky:2009dc,
Oliveira:2008uf,Oliveira:2009eh}, leading  
not only to quantitative refinements, but, in some cases, to critical revisions  
of the underlying physical concepts~\cite{Aguilar:2008xm,Boucaud:2008ji,Pennington:2011xs,Bashir:2012fs}.

The quantitative understanding of the ghost sector of QCD 
constitutes a long-standing challenge for the SDE practitioners. 
Without a doubt, the most fundamental quantity in this context 
is the ghost propagator, $D(p^2)$, and the corresponding dressing function, 
$F(p^2) = p^2 D(p^2)$; in fact, the 
infrared behavior of the latter, in the Landau gauge (LG), has been traditionally   
associated with a particular realization of color confinement~\cite{Kugo:1979gm,Kugo:1995km,Watson:2001yv,Kondo:2011ab}. 

In recent years, 
various lattice studies, both in $SU(2)$
and  $SU(3)$, together with numerous analytic approaches, 
find a massless ghost propagator with an infrared finite dressing function~\cite{Cucchieri:2007md,Bogolubsky:2007ud,Aguilar:2008xm,Boucaud:2008ji,Dudal:2008sp}.
In addition, in the same gauge, 
the gluon propagator obtained on the lattice is finite in the deep infrared, 
supporting the notion of an effectively massive gluon.  
In fact, the dynamical gluon mass generation, 
first proposed in~\cite{Cornwall:1981zr}, and further developed in a number of recent works,
provides a unified explanation for the observed finiteness of both aforementioned 
quantities~\cite{Aguilar:2004sw, Aguilar:2006gr,Aguilar:2008xm, Binosi:2007pi, Aguilar:2011yb}. Specifically, 
an infrared finite $F(p^2)$ emerges as a direct consequence 
of the massiveness of the gluon propagator: 
such a gluon propagator, when inserted 
in the SDE of the ghost propagator, saturates the logarithms 
associated with the $F(p^2)$, thus making it finite at the origin.

However, 
what has been more difficult to obtain from a self-consistent SDE analysis  
is the entire shape and size of $F(p^2)$ provided by the lattice~\cite{Aguilar:2008xm,Dudal:2012zx}.
In fact, even when one substitutes into the 
ghost SDE the gluon propagator 
furnished by the lattice, but keeping the  ghost-gluon  vertex at its tree-level value, 
the resulting $F(p^2)$ is significantly suppressed compared to that of the lattice~\cite{Aguilar:2008xm};
to reproduce the lattice result, one has to artificially 
increase the value of the gauge coupling 
from the correct value $\alpha_s = 0.22$ to $\alpha_s = 0.29$~\cite{Aguilar:2010gm}. 

It would seem, therefore, that  
the main reason for the observed discrepancy ought to be traced back to the 
way in which the fully dressed ghost-gluon vertex, $\Gamma_{\nu}$, 
appearing in the 
ghost SDE, is approximated.
Even though preliminary lattice studies indicate that 
the deviations of $\Gamma_{\nu}$ from 
its tree-level value are relatively moderate~\cite{Cucchieri:2004sq,Ilgenfritz:2006gp,Ilgenfritz:2006he,Sternbeck:2006rd,Cucchieri:2008qm},  
the highly non-linear nature of 
the ghost SDE may lead to considerable enhancements. 
In fact, a modest increase of the relevant vertex form factor 
in the region of momenta that provide the largest support to the 
ghost SDE may account for the bulk of the required effect.

The purpose of this article is to obtain 
a reliable ``first-principle'' approximation for 
this important missing ingredient.
Specifically, we will 
determine the relevant vertex form factor 
from an approximate version of the SDE satisfied by the vertex $\Gamma_{\nu}$ itself, in the LG. 
To be sure, the vertex SDE has a complicated skeleton expansion, involving various 
unknown (or only partially known) quantities, such as multiparticle kernels. 
The basic approximations we employ at the level of the vertex SDE are: 
{\it (i)} we consider only the first two diagrams in this expansion; this corresponds to the 
``one-loop dressed'' truncation~\cite{Schleifenbaum:2004id}, and {\it (ii)} inside these diagrams 
we replace full vertices by their 
tree-level values, but keep {\it fully dressed} ghost and gluon propagators, 
(iii) for the numerical analysis of the resulting integrals, we use 
as input for the full gluon  propagators the  lattice data of~\cite{Bogolubsky:2007ud}.

The tensorial decomposition of  $\Gamma_{\nu}$ consists of two form factors [see~\1eq{Gtens}];
however, given that this vertex will be inserted 
in the ghost SDE, written in the  LG, only the cofactor $A(-k,-p,r)$ 
of the ghost momentum $p_{\nu}$ survives.
In the present study we determine $A(-k,-p,r)$ for two particular kinematic 
 configurations, soft gluon ($k \to 0$) and soft ghost ($p \to 0$), thus converting it, in both cases,  
to a function of a single momentum only, $A(0,-p,p)$ and $A(-k,0,k)$, respectively.
In fact, as we will explain in detail in Sec.~\ref{sec.vertexSDE}, 
the case where $p\to 0$ is equivalent to the standard Taylor limit~\cite{Taylor:1971ff,Marciano:1977su}.

Our main results may be summarized as follows. 

({\it i}) In the soft gluon limit, the result obtained for $A(0,-p,p)$ 
displays a moderate peak around 1 GeV, corresponding to a 20$\%$ increase 
with respect to the tree-level value; this result   
compares rather well with the existing lattice data~\cite{Ilgenfritz:2006he,Sternbeck:2006rd}. 
Of course, this particular kinematic configuration is not relevant for the ghost SDE, but serves 
as a preliminary test of the overall faithfulness of the approximations employed.

({\it ii}) The numerical solutions for the 
coupled system of integral equations determining $F(p^2)$ and $A(-k,0,k)$
gives rise to a ghost dressing function that is in excellent agreement with the 
lattice data~\cite{Bogolubsky:2007ud}. The corresponding solution for $A(-k,0,k)$ is characterized by a 
rather pronounced maximum, centered again around  \mbox{$1$ GeV}, reaching a value of 
about $1.5$. 
In this analysis we use \mbox{$\alpha_s = 0.22$}, which corresponds to 
the momentum-subtraction (MOM) value for 
the point \mbox{$\mu=4.3$ GeV}~\cite{Boucaud:2008gn}, used 
to renormalize the gluon propagator obtained from the lattice.

The article is organized as follows. 
In Sec.~\ref{Sec.def} we introduce the necessary notation, and set up the 
SDE for the ghost dressing function, paying particular attention to 
the way that the fully dressed ghost-gluon vertex enters in it. 
In Sec.~\ref{sec.vertexSDE} we carry out the analysis at the level of the 
SDE of the ghost-gluon vertex, and derive the corresponding closed expressions in the 
two kinematic limits of interest. In Sec.~\ref{numsec} we present     
the numerical treatment of the equations derived in the previous sections. In particular, we first 
compute the case of the soft gluon, and then we proceed to the solution of the coupled system.  
Finally, our  conclusions and discussion are presented in Sec.~\ref{concl}.

\section{Ghost dressing function and the ghost-gluon vertex}\label{Sec.def}

\begin{figure}[b]
\includegraphics[scale=0.85]{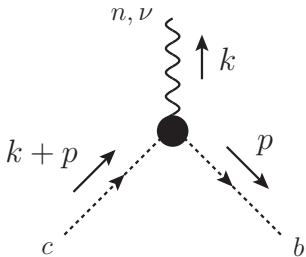}
\caption{The fully dressed ghost-gluon vertex.}
\label{fullgg}
\end{figure}

In this section we introduce the
SDE for the ghost propagator in the LG,
and discuss some of its basic properties and features. 
Of particular interest is the dependence of this equation on 
the surviving component of the ghost-gluon vertex, 
and the numerical implications of approximating it by its tree-level value. 

Our starting point is the full ghost-gluon vertex,
shown in Fig.~\ref{fullgg}, and denoted by 
\bea
\Gamma_{\nu}^{nbc}(-k,-p,r) = gf^{nbc}\Gamma_{\nu}(-k,-p,r)\,, \quad r=k+p\,,
\label{Gdef}
\eea
with $k$ representing the momentum of the gluon and $p$ of the anti-ghost. 
The   most general tensorial structure of this vertex is given by
\be
{\Gamma}_{\nu}(-k,-p,r) =  A(-k,-p,r)\, p_{\nu} + B(-k,-p,r) \,k_{\nu} \,;
\label{Gtens}
\ee 
at tree-level, the two form factors assume the values 
\mbox{$A^{[0]}(-k,-p,r)=1$} and $B^{[0]}(-k,-p,r)=0$,  giving rise to the bare ghost-gluon vertex  
$\Gamma^{[0]}_{\nu}=p_\nu$.

The form factors $A$ and $B$ may be formally projected out by contracting ${\Gamma}_{\nu}$  
with the vectors
\be
\varepsilon_{\nu}^{\chic A}(k,p) = \frac{k^2 p_{\nu} -(k\cdot p)k_{\nu}}{k^2p^2 -(k \cdot p)^2}
\,,\qquad
\varepsilon_{\nu}^{\chic B}(k,p) = \frac{p^2 k_{\nu} -(k \cdot p)p_{\nu}}{k^2p^2 -(k \cdot p)^2}\,,
\label{proj1}
\ee
namely
\be
A(-k,-p,r)= \varepsilon_{\nu}^{\chic A}(k,p) {\Gamma}^{\nu}(-k,-p,r)
\,,\hspace{1.cm}
B(-k,-p,r) = \varepsilon_{\nu}^{\chic B}(k,p) {\Gamma}^{\nu}(-k,-p,r)\,.
\label{proj2}
\ee

Of particular importance for the analysis that follows is 
the so-called ``Taylor limit'' of the ghost-gluon vertex, corresponding to the case of vanishing ghost momentum, $r=0$, $p=-k$. 
In this special kinematic configuration, the $\Gamma_\nu (-k,-p,r)$ of Eq.~(\ref{Gtens}) becomes
\be
\Gamma_\nu(-k,k,0) = -[A(-k,k,0) - B(-k,k,0)]k_\nu \,.
\label{GtensTayl}
\ee
Closely related to this limit is the well-known Taylor theorem, which states that, to all orders in perturbation theory,
\be
A(-k,k,0) - B(-k,k,0) = 1;
\label{Taylteor}
\ee
as a result, the fully-dressed vertex assumes the tree-level value corresponding to this 
 particular kinematic configuration, \ie $\Gamma_\nu(-k,k,0)=-k_\nu$.

\begin{figure}[!t]
\includegraphics[scale=0.7]{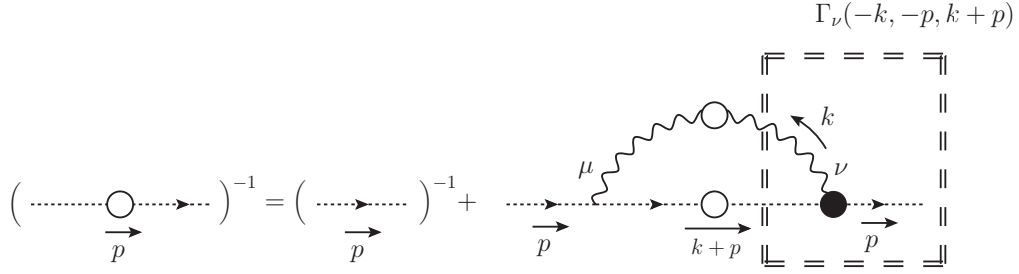}
\caption{The SDE for the ghost propagator given by Eq.~(\ref{SDgh}). The white blobs represent the
fully dressed gluon and ghost propagators, while the black blob denotes the dressed ghost-gluon
vertex.}
\label{ghostSDE}
\end{figure}

After these introductory comments, let us 
turn to the SDE for the ghost propagator, and examine in some detail 
how the ghost-gluon vertex affects its structure. 
The relevant SDE is diagrammatically represented in the Fig.~\ref{ghostSDE}. Using
the momenta flow and Lorentz indices as indicated in Fig.~\ref{ghostSDE}, the ghost SDE  can be written as
\be
iD^{-1}(p^2) = ip^2 - g^2 C_{\rm {A}}  \int_k
\Gamma_{\mu}^{[0]}(k,-k-p,p)\Delta^{\mu\nu}(k)\Gamma_{\nu}(-k,-p,k+p) D(k+p)\,,
\label{SDgh}
\ee
where $C_{\rm A}$ denotes the Casimir eigenvalue of the adjoint representation ($N$ for $SU(N)$), 
$d=4-\epsilon$ is the space-time dimension, and we have introduced the integral measure
\be
\int_{k}\equiv\frac{\mu^{\epsilon}}{(2\pi)^{d}}\!\int\!\mathrm{d}^d k,
\label{dqd}
\ee
with $\mu$ the 't Hooft mass.
In the LG, the gluon
propagator $\Delta_{\mu\nu}(q)$ has the transverse form
\be 
\Delta_{\mu\nu}(q)=-i P_{\mu\nu}(q)\Delta(q^2)\,,
\label{prop_cov}
\ee
with  
\be
P_{\mu\nu}(q)= g_{\mu\nu} - \frac{q_\mu q_\nu}{q^2} \,, 
\ee
the usual projection operator. 

Clearly, due to the full transversality of $\Delta_{\mu\nu}(k)$, 
any reference to the form factor $B$  
disappears from the ghost SDE of \1eq{SDgh}. Specifically, 
substituting \1eq{Gtens} into \1eq{SDgh} we obtain 
\be
F^{-1}(p^2) = 1 +ig^2 C_{\rm {A}} \int_k\, \left[1-\frac{(k\cdot p)^2}{k^2p^2}\right] A(-k,-p,k+p)\Delta (k)  D(k+p) \,,
\label{tt2}
\ee
where we have  introduced the ghost dressing function, $F(q^2)$, defined as
\be
D(q^2)= \frac{F(q^2)}{q^2}\,.
\label{ghostdress}
\ee

The renormalization of \1eq{tt2} proceeds through the replacements 
\bea
\Delta_{\chic R}(q^2)&=& Z^{-1}_{A} \Delta(q^2),\nonumber\\
F_{\chic R}(q^2)&=& Z^{-1}_{c} F(q^2),\nonumber\\ 
\Gamma^{\nu}_{\chic R}(q,p,r) &=& Z_1 \Gamma^{\nu}(q,p,r),\nonumber\\ 
g_{\chic R} &=& Z_g^{-1} g = Z_1^{-1} Z_A^{1/2} Z_c\, g \,,
\label{renconst}
\eea
where  $Z_{A}$, $Z_{c}$, $Z_{1}$, and $Z_g$ are the corresponding
renormalization constants; the dependence of the above quantities on the renormalization point $\mu$ 
is suppressed. 
In the MOM scheme, 
usually employed in the SDE analysis, 
the renormalization conditions imposed are that at $\mu$
the corresponding Green's functions assume  
their tree-level values, e.g.,  \mbox{$\Delta_{\chic R}^{-1}(q^2=\mu^2)= \mu^2$}, and  
$F_{\chic R}(q^2=\mu^2) =1$~\cite{Marciano:1977su}.
Note also that, in the LG, the form factor $A$  is ultraviolet finite at one-loop, 
and therefore, no infinite renormalization constant needs to be introduced at that order for $\Gamma^{\nu}$.
In fact, one usually invokes Taylor's theorem [see \1eq{Taylteor}],
in order to finally set $Z_1 =1$ to all orders (see discussion in Sec.~\ref{sec.vertexSDE}). 

Then, the SDE becomes 
\be
F^{-1}(p^2) = Z_c +ig^2 C_{\rm {A}} \int_k\, \left[1-\frac{(k\cdot p)^2}{k^2p^2}\right] A(-k,-p,k+p)\Delta (k)  D(k+p) \,,
\label{tt2r}
\ee
where we have suppressed the subscript ``R'' to avoid notation clutter. 
The actual closed expression of $Z_c$ is obtained from \1eq{tt2r} itself, by imposing the aforementioned 
MOM renormalization condition on $F^{-1}(p^2)$.

Evidently, 
the explicit dependence of \1eq{tt2r} on $A(-k,-p,k+p)$ requires the use of the corresponding vertex SDE, 
thus converting the problem of determining $F(p^2)$ into a coupled SDE system.  
The usual way to circumvent this technical complication has been to simply approximate $A(-k,-p,k+p)$ 
by its tree-level value, setting into \1eq{tt2r} \mbox{$A(-k,-p,k+p) =1$}.

\begin{figure}[!t]
\includegraphics[scale=0.55]{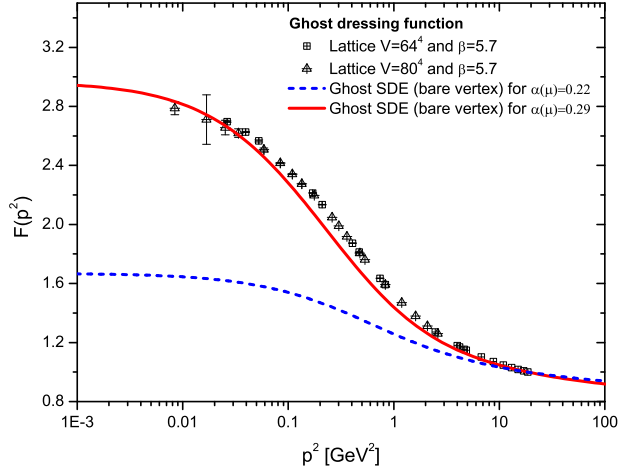}
\caption{Comparison of the ghost dressing function, $F(p^2)$, obtained as  
solution of the ghost SDE when the ghost-gluon vertex is approximate by its bare value, with the
lattice data of Ref.~\cite{Bogolubsky:2007ud}.
The (red) continuous curve represents the case when  $\alpha_s(4.3 \rm GeV)=0.29$ whereas
the (blue) dotted curve is obtained when  $\alpha_s( 4.3 \rm GeV )=0.22$.}
\label{oldsol}
\end{figure}

Then, after proper renormalization along the lines discussed above,
and passing to the Euclidean space following the standard rules, one 
solves  \1eq{tt2r} numerically, using 
the lattice data of~\cite{Bogolubsky:2007ud} as input for the 
gluon propagator. Note that this latter propagator is renormalized within the MOM scheme, 
by imposing the standard condition $\Delta^{-1}(\mu^2)=\mu^2$ at  \mbox{$\mu=4.3$ GeV}, 
namely the deepest available point in this set of lattice data; then,  
the corresponding 
value for $\alpha_s= g^2/4\pi$ that one should use is \mbox{$\alpha_s(4.3 \rm GeV)=0.22$}. 
However, for this particular value of $\alpha_s$, the solution obtained from \1eq{tt2r}
lies considerably below the lattice data for $F(p^2)$,
as can be clearly seen from the (blue) dotted  curve of Fig.~\ref{oldsol}.
In order to obtain a close coincidence with the lattice, one must increase the value of $\alpha_s(4.3 \rm GeV)$ to 0.29, 
thus obtaining the (red) continuous curve in Fig.~\ref{oldsol}.

It is, of course, natural to attribute the observed discrepancy to the  aforementioned simple   
approximation employed for the ghost-gluon vertex. Therefore, to ameliorate the 
situation, we will determine this form factor from its corresponding SDE, in a certain  
kinematic limit that is relevant for the situation at hand.
Specifically, given that $A(-k,-p,k+p)$ is a function of three variables, 
$p^2$, $k^2$, and the angle between the two (appearing in the inner product $p\cdot k$), 
a full SDE treatment is rather cumbersome, and lies beyond our present technical powers.
Instead, we will consider the behavior of $A(-k,-p,k+p)$ for vanishing $p$; to that end, 
we start out with the Taylor expansion of $A(-k,-p,k+p)$ around $p=0$,  
and we only keep the first term, $A(-k,0,k)$, thus converting $A$ into a function of a single variable.

We emphasize that the limit $p\to 0$ is taken only inside the argument of the form factor $A$, but not in the rest 
of the terms appearing in the SDE of \1eq{tt2r}.
Specifically, following the procedure explained in detail in the next section, 
one isolates from the ghost-gluon SDE the contribution proportional to $p_{\nu}$, 
taking the limit $p\to 0$ in the accompanying scalar co-factor, thus arriving at a form 
$\Gamma_{\nu}(-k,-p,k+p) = p_{\nu} A (-k,0,k)$. 
Equivalently, in terms of the projectors introduced in Eqs.~(\ref{proj1}) and (\ref{proj2}), one has
\be
A(-k,0,k) = \lim_{p\rightarrow 0}\left\{ \varepsilon_{\nu}^{\chic A}(k,p) {\Gamma}^{\nu}(-k,-p,k+p)\right\}\,.
\ee 
Thus, the approximate version of the SDE in \1eq{tt2r} reads
\be
F^{-1}(p^2) = Z_c +ig^2 C_{\rm {A}} \int_k\, \left[1-\frac{(k\cdot p)^2}{k^2p^2}\right] A(-k,0,k)\Delta (k)  D(k+p) \,.
\label{tt2app}
\ee


\section{The ghost-gluon vertex}\label{sec.vertexSDE}

In this section we derive in detail the nonperturbative expression for the form factor $A$,  
in two special kinematic configurations: (i) the \emph{soft gluon limit}, in which the momentum 
carried by the gluon leg is zero ($k=0$), and (ii) the \emph{soft ghost limit}, 
where the momentum of the anti-ghost leg vanishes ($p=0$). 

\subsection{General considerations}

\begin{figure}[!t]
\includegraphics[width=16cm]{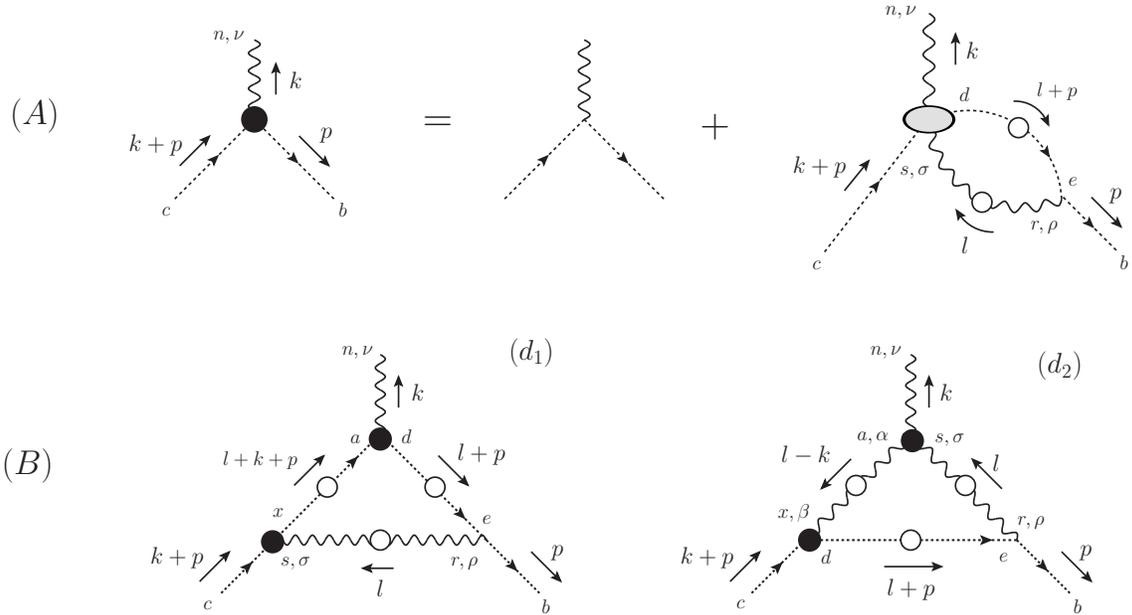}
\caption{(A) The complete SDE of the ghost-gluon vertex. Notice that we have set up it with respect to the anti-ghost leg. (B) Diagrams included in the skeleton expansion of the ghost-gluon kernel that we will consider in our analysis.}
\label{vertexSDE}
\end{figure}

The starting point 
of our analysis is the SDE satisfied by the ghost-gluon vertex, whose 
diagrammatic representation is shown in panel $(A)$ of Fig.~\ref{vertexSDE}. One observes  
that the relevant quantity, which controls the dynamics of this SDE, is the four-point ghost-gluon kernel. 

For the ensuing analysis we will carry out the following main simplifications:

({\bf i}) The ghost-gluon kernel will be replaced by its ``one-loop dressed'' approximation;    
specifically, in the corresponding skeleton expansion we will only include 
the diagrams appearing in panel $(B)$ of Fig.~\ref{vertexSDE}. 
Thus, the approximate version 
of the SDE that we employ may be cast in the form
\begin{equation}\label{glghSDE}
\Gamma_\nu(-k,-p,k+p) = p_\nu - \frac{i}{2}g^2 C_{\rm A} [(d_1)_\nu - (d_2)_\nu]\,,
\end{equation}
where the diagrams $(d_i)$ are given by
\begin{eqnarray}\label{SDEdiagrams}
(d_1)_\nu &=& \int_l \Gamma_\rho^{[0]}\Delta^{\rho\sigma}(l)\Gamma_\sigma D(l+k+p) \Gamma_\nu D(l+p)\,, \nonumber \\
(d_2)_\nu &=& \int_l \Gamma_\rho^{[0]}\Delta^{\rho\sigma}(l)\Gamma_{\nu\sigma\alpha}\Delta^{\alpha\beta}(l-k)\Gamma_\beta D(l+p)\,.
\end{eqnarray}
For notational simplicity, we have suppressed the arguments of the momenta in all vertices; the latter may be 
easily recovered from the figures and the conventions established in Sec.~\ref{Sec.def}. Note also 
that, in the LG that we use, 
the gluon propagators appearing in the above expressions assume the completely transverse form of \1eq{prop_cov}.

({\bf ii})
The (multiplicative) renormalization of \1eq{glghSDE} proceeds in the standard way.
Specifically, in addition 
to the renormalization constants and relations given in \1eq{renconst}, one 
must introduce the vertex renormalization for the three-gluon vertex, to be denoted by $Z_3$, namely 
$\Gamma^{\nu\sigma\alpha}_{\chic R} =  Z_3 \Gamma^{\nu\sigma\alpha}$, together with the 
corresponding relation for the coupling renormalization, namely 
$g_{\chic R} = Z_3^{-1} Z_A^{3/2} \, g$. From this relation, and the last of \1eq{renconst}, 
one has that   $Z_3^{-1} Z_A = Z_1^{-1} Z_c$. Then, it is straightforward to show that the 
contributions of $g^2(d_1)_\nu$ and $g^2(d_1)_\nu$   
maintain the same form after renormalization; in fact, this property may be easily 
established by grouping the integrands  
in terms of the standard renormalization-group invariant quantities formed 
by $(g \Gamma_{\mu} \Delta^{1/2} D)$ and $(g \Gamma^{\nu\sigma\alpha} \Delta^{3/2})$~\cite{Aguilar:2009nf}. Thus,  
the renormalized version of \1eq{glghSDE} reads 
\be
\Gamma_{\chic R}^\nu(-k,-p,k+p) = Z_1 \left\{ 
p_\nu - \frac{i}{2} g_{\chic R}^2 C_{\rm A} [(d_1)_{\chic R}^\nu - (d_2)_{\chic R}^\nu ]\right\} \,,
\label{glghSDEren}
\ee
where the $Z_1$ comes from the renormalization of the $\Gamma^\nu(-k,-p,k+p)$ on the lhs. 

In what follows we will set $Z_1=1$. 
In the case of the soft ghost configuration, $p=0$, (which, as we will see, is equivalent to the Taylor kinematics), 
this choice is imposed by Taylor's theorem, see \1eq{Taylteor}. 
On the other hand, 
in the case of the soft gluon configuration, $k=0$, this choice constitutes 
an approximation, 
in the sense that it is motivated by the one-loop finiteness of the (LG) $\Gamma^\nu$, 
but is not enforced by an analogous all-order relation.

({\bf iii}) In the two aforementioned diagrams, $(d_1)$ and $(d_2)$,  we will keep fully dressed propagators, but will 
replace the fully dressed three-gluon vertex appearing in graph $(d_2)$ 
by the 
corresponding tree-level expression, namely
\begin{equation}
\label{treeglvertex}
\Gamma_{\alpha\mu\nu}(q,r,p) \rightarrow
\Gamma_{\alpha\mu\nu}^{[0]}(q,r,p) = (r-p)_\alpha g_{\mu\nu} + (p-q)_\mu g_{\nu\alpha} + (q-r)_\nu g_{\alpha\mu}\,.
\end{equation}
Furthermore, as will be explained in the corresponding subsections,  
additional approximations will be imposed on the fully dressed ghost-gluon vertices, 
depending on the specific details of each kinematic case considered.

\subsection{\label{softh}Soft gluon configuration}

We begin with the analysis of the soft gluon configuration, $k=0$. 
Evidently, in this case the ghost-gluon vertex becomes a function of only one momentum, $p$, 
and may be described in terms of a single form factor, namely,
\begin{equation}\label{vertexA}
\Gamma_\nu(0,-p,p) = A(p) p_\nu; \quad A(p)\equiv A(0,-p,p)\,.
\end{equation}
Therefore, setting $k=0$ in Eq.~(\ref{glghSDE}), one is able to isolate the form factor $A$ by means of the projection
\begin{equation}\label{factorA}
A(p) = 1 - \frac{i}{2}g^2 C_A [(d_1) - (d_2)]; \quad (d_i) \equiv \frac{p^\nu}{p^2}(d_i)_\nu\,, \quad i=1,2\,,
\end{equation}
where the diagrams $(d_i)$ are obtained from those of Eq.~(\ref{SDEdiagrams}) in the limit $k\rightarrow 0$.

The particular kinematic configuration considered here 
allows one to derive 
a linear integral equation for the unknown quantity  $A(p)$. 
This becomes possible because, in the limit $k=0$, 
the 
vertex $\Gamma_{\nu}$ entering in graph $(d_1)$ becomes $\Gamma_{\nu}(0, -l-p,l+p)$.
Thus, the integral $(d_1)$ contains $A(0, -l-p,l+p)$,  
giving rise to an integral equation for $A(0,-p,p)$.
Unfortunately, this favorable set of circumstances 
does not apply to the remaining 
ghost-gluon vertices, namely, $\Gamma_\sigma$ and $\Gamma_\beta$ in graphs $(d_1)$ and $(d_2)$, 
respectively; their arguments depend on all possible 
kinematic variables, and the inclusion of the full $A$ 
would give rise to a (non-linear) integral equation, too complicated to solve. 
We therefore approximate all remaining ghost-gluon vertices by their tree-level 
expressions.

After these comments, and use of the notation introduced in \1eq{factorA}, the 
diagram $(d_1)$ reads 
\begin{equation}\label{d1projection}
(d_1) = \int_l \frac{(l\cdot p)}{(l+p)^2 p^2}[(l\cdot p)^2 - l^2 p^2] D^2(l)\Delta(l+p)A(l)\,.
\end{equation}

To evaluate the contribution of diagram $(d_2)$ notice that, with the gluon propagators in the LG, 
and the bare three-gluon vertex of Eq.~(\ref{treeglvertex}), we have that
\begin{equation}\label{projectorsthreegl}
P^{\rho\sigma}(l)P^{\alpha\beta}(l)\Gamma^{[0]}_{\nu\sigma\alpha}(0,l,-l) = 2l_\nu P^{\rho\beta}(l)\,.
\end{equation}
Applying this result we get
\begin{equation}\label{d2projection}
(d_2) = 2 \int_l \frac{(l\cdot p)}{l^2 p^2}[l^2 p^2 - (l\cdot p)^2] \Delta^2(l) D(l+p)\,.
\end{equation}

The final answer is obtained by substituting 
\1eq{d1projection} and \1eq{d2projection} into \1eq{factorA}; it will be written directly in
Euclidean space, using the standard  transformation rules,  
\be
-q^2=q^2_\mathrm{\chic E}\,; \quad \Delta_\mathrm{E}(q^2_\mathrm{\chic E})=-\Delta(-q^2_\mathrm{\chic E})\,;\quad D_\mathrm{\chic E}(q^2_\mathrm{\chic E})=-D(-q^2_\mathrm{\chic E})\,; \quad
\int_k=i\int_{k_\mathrm{\chic E}}\,,  
\label{Euclidean}  
\ee
and setting 
\begin{eqnarray}\label{spherical}
&& l^2=t\,;\quad p^2=x\,;\quad (l+p)^2=z\,;\quad (l\cdot p) = \sqrt{xt}\cos\theta\,; \nonumber \\
&& \int_{l_E}=\int\frac{d^4l}{(2\pi)^4} = \frac{1}{(2\pi)^3}\int_0^\infty \!\!\! dt\,t\int_0^\pi \!\!\! d\theta \sin^2\theta\,.
\end{eqnarray}
Specifically (we suppress the subscript ``E''), 
\begin{eqnarray}\label{Aeuclidean}
A(x) &=& 1 - \frac{\alpha_s C_A}{4\pi^2}\int_0^\infty \!\!\! dt \sqrt{xt}\,F^2(t) A(t) \int_0^\pi \!\!\!d\theta\sin^4\theta\cos\theta\left[\frac{\Delta(z)}{z}\right] \nonumber \\
&-& \frac{\alpha_s C_A}{2\pi^2}\int_0^\infty \!\!\! dt \sqrt{xt}\, t\,\Delta^2(t) \int_0^\pi \!\!\! d\theta\sin^4\theta\cos\theta\left[\frac{F(z)}{z}\right] \,,
\end{eqnarray}
where we have used $g^2=4\pi \alpha_s$, and Eq.~(\ref{ghostdress}) 
in order to express the ghost propagators in terms of their dressing functions. 

Notice that, in the limit $x=0$, namely when the momentum of the ghost leg is also zero, 
we recover from Eq.~(\ref{Aeuclidean}) the tree-level value of the form factor, \textit{i.e.}, $A(0)=1$.

\subsection{Soft ghost configuration (Taylor kinematics)}

We next turn to the case that, according to the discussion presented in Sec.~\ref{Sec.def},  is expected to 
improve the treatment of the ghost SDE. Specifically,  
in this subsection we will derive an approximate version for $A$ in the soft ghost configuration, to be denoted by
\begin{equation}\label{Asoftghost}
\lim_{p\rightarrow 0}A(-k,-p,k+p) = A(-k,0,k) \equiv A(k)\,.
\end{equation}

However, before proceeding to this derivation, 
we will demonstrate that the form factor $A(-k,0,k)$ 
obtained in the soft ghost configuration 
is none other than the form factor $A(-k,k,0)$, appearing  
in the constraint imposed by Taylor's theorem, given by Eq.~(\ref{Taylteor}). 
To prove that, let us rewrite the SDE of the ghost propagator, Eq.~(\ref{SDgh}),  
dressing this time the left ghost-gluon vertex instead of the right, {\it i.e.},
\be
iD^{-1}(p^2) = ip^2 - g^2 C_{\rm {A}}  \int_k
\Gamma_\mu(k,-k-p,p)\Delta^{\mu\nu}(k)\Gamma_\nu^{[0]}(-k,-p,k+p) D(k+p)\,,
\label{leftSDE}
\ee
where we have maintained the same momenta flow and Lorentz indices as in Fig.~\ref{ghostSDE}. 
Therefore, using Eq.~(\ref{Gtens}) for the $\Gamma_\mu$ in Eq.~(\ref{leftSDE}), we get (in the LG)
\be
F^{-1}(p^2) = 1 +ig^2 C_{\rm {A}}\int_k\, \left[1-\frac{(k\cdot p)^2}{k^2p^2}\right] A(k,-k-p,p) \Delta (k) D(k+p).
\label{leftdressing}
\ee
Evidently Eqs.~(\ref{tt2}) and (\ref{leftdressing}) must furnish an identical result 
for $F(p^2)$, since the answer cannot depend on which of the two vertices one chooses to dress. 
Thus, the form factor $A$ is forced to satisfy the equality
\be
A(-k,-p,k+p) = A(k,-k-p,p).
\label{Asymone}
\ee
Given that, due to Lorentz invariance, the dependence on the momenta is quadratic, {\it i.e.}, 
$A(k^2,p^2, k^2+p^2 + 2 k\cdot p)$, we have immediately that 
\be
A(k,-k-p,p) = A(-k,k+p,-p).
\label{Asymtwo}
\ee
So, combining Eqs.~(\ref{Asymone}) and (\ref{Asymtwo}), we arrive at the relation 
\be
A(-k,-p,k+p) = A(-k,k+p,-p),
\label{Asymthree}
\ee
which states that, in the LG, the form factor $A$ of the gluon-ghost vertex is invariant under 
the exchange of the momenta of the ghost and anti-ghost legs. Notice that this invariance 
is known to be a consequence of a global $SL(2,R)$ symmetry between the ghost 
and anti-ghost fields~\cite{Alkofer:2000wg}, which implies that the LG  
is a ghost--anti-ghost symmetric gauge fixing choice. 
Finally, setting $p=0$ in Eq.~(\ref{Asymthree}), 
one obtains the announced result, that is, the $A$ obtained in the soft ghost limit coincides with that of the Taylor kinematics. 

As mentioned above,
the fact that the kinematic situation considered here is equivalent to the Taylor limit, imposes, in a natural way, 
the value $Z_1=1$ for the renormalization constant appearing in \1eq{glghSDEren}.

Once the above connection has been established, we return to the derivation 
of the explicit expression for the form factor $A$ in the soft ghost limit. 
To that end, we will consider again the diagrams shown in panel $(B)$ of Fig.~\ref{vertexSDE},  
with dressed gluon and ghost propagators, and tree-level values for \textit{all} the interaction vertices. 
In this configuration, the expressions given in Eq.~(\ref{SDEdiagrams}) reduce to
\begin{eqnarray}\label{diagramsp0}
(d_1)_\nu &=& p^\rho (k+p)^\sigma \int_l (l+p)_\nu D(l+p)D(l+k+p)\Delta(l)P_{\rho\sigma}(l)\,, \nonumber \\
(d_2)_\nu &=& p^\rho (k+p)^\beta \int_l D(l+p)\Delta(l)\Delta(l-k)P_\rho^\sigma(l)P^\alpha_\beta(l-k)\Gamma^{[0]}_{\nu\sigma\alpha}\,,
\end{eqnarray}

We next outline the general procedure for isolating the $A(-k,0,k)$ defined in \1eq{Asoftghost}. 
First, we observe that the most general Lorentz decomposition of the diagrams given in Eq.~(\ref{diagramsp0}) is
\begin{eqnarray}\label{generalexp}
(d_i)_\nu &=& p^\rho (k+p)^\sigma [f_1 g_{\nu\rho} k_\sigma + f_2 g_{\nu\sigma} k_\rho + f_3 g_{\rho\sigma} k_\nu + f_4 g_{\nu\rho} p_\sigma + f_5 g_{\nu\sigma} p_\rho + f_6 g_{\rho\sigma} p_\nu \nonumber \\
&+& f_7 p_\nu p_\rho p_\sigma + f_8 p_\nu p_\rho k_\sigma + f_9 p_\nu k_\rho p_\sigma + f_{10} p_\nu k_\rho k_\sigma + f_{11} k_\nu k_\rho k_\sigma + f_{12} k_\nu k_\rho p_\sigma \nonumber \\
&+& f_{13} k_\nu p_\rho k_\sigma + f_{14} k_\nu p_\rho p_\sigma]\,,
\end{eqnarray}
where the corresponding form factors $f_i\equiv f_i(k,p)$ are assumed to be finite 
in the infrared limit $p\rightarrow 0$. 

A detailed look at this expansion reveals that only the tensorial structure $g_{\nu\rho} k_\sigma$, accompanying the form factor $f_1$, can saturate the prefactor $p^\rho (k+p)^\sigma$ and survive when the limit $p\rightarrow 0$ is taken. 
Specifically, we may rewrite Eq.~(\ref{generalexp}) as follows
\begin{eqnarray}\label{p0survive}
(d_i)_\nu &=& p^\rho k^\sigma f_1(k,p)g_{\nu\rho} k_\sigma + {\cal O}(p)(k+p)_\nu \nonumber \\
&=& k^2f_1(k,p)p_\nu + {\cal O}(p)(k+p)_\nu\,,
\end{eqnarray}
where the symbol ${\cal O}(p)(k+p)_\nu$ is used to indicate terms that saturate with $p_\nu$ or $k_\nu$, but whose form factors are of order ${\cal O}(p)$ or higher, 
and will not contribute in the soft ghost configuration. 
Furthermore, one can perform the Taylor expansion of $f_1(k,p)$ around $p=0$, namely,
\begin{equation}\label{Taylorf1}
f_1(k,p) = f_1(k,0) + 2(k\cdot p) f_1'(k) + {\cal O}(p^2); \quad f_1'(k) \equiv \frac{\partial}{\partial p^2}f_1(k,p)\bigg\vert_{p=0}\,.
\end{equation}
Thus, only the zero order term of this expansion is relevant for our kinematic configuration, and we obtain finally from Eq.~(\ref{p0survive}) the following result
\begin{equation}\label{p0term}
(d_i)_\nu = k^2f_1(k,0)p_\nu + {\cal O}(p)(k+p)_\nu\,,
\end{equation}
where the quantity $k^2f_1(k,0)$ should be identified as the contribution of the corresponding diagram to $A(k)$, while   
terms containing the derivatives of $f_1$ are naturally reassigned to ${\cal O}(p)(k+p)_\nu$. 

After these observations, it is relatively easy to establish 
that this generic procedure can be systematically implemented by performing the following steps: ({\it i}) Set $p=0$ 
from the beginning \textit{inside} the integrals of Eq.~(\ref{diagramsp0}). ({\it ii}) Discard all the terms that give rise to structures 
of the type ${\cal O}(p)(k+p)_\nu$. 
({\it iii}) Determine the contribution of the diagram that saturates the index of the momentum $p^\rho$ with the metric tensor $g_{\nu\rho}$.

To illustrate in some detail the above procedure, 
let us focus our attention on the contribution of diagram $(d_1)$, appearing in the first line of Eq.~(\ref{diagramsp0}). Applying step ({\it i}), we obtain
\begin{equation}\label{d1i}
(d_1)_\nu = p^\rho(k+p)^\sigma \bigg\lbrace \int_l l_\nu D(l) D(l+k)\Delta(l)P_{\rho\sigma}(l) + p_\nu \int_l D(l) D(l+k)\Delta(l)P_{\rho\sigma}(l) \bigg\rbrace\,.
\end{equation}
Now, using criterion ({\it ii}), it is easy to recognize that the part of Eq.~(\ref{d1i}) to be retained is given by
\begin{equation}\label{d1ii}
(d_1)_\nu = -p^\rho I_{\nu\rho}(k)\,,
\end{equation}
where we have defined the integral
\begin{equation}\label{intI}
I_{\nu\rho}(k) = \int_l \frac{(l\cdot k)}{l^2}D(l)D(l+k)\Delta(l) l_\nu l_\rho\,,
\end{equation}
which may be further decomposed as
\begin{equation}\label{Idecom}
I_{\nu\rho}(k) = I_1(k^2) g_{\nu\rho} + I_2(k^2)k_\nu k_\rho\,,
\end{equation}
with 
\begin{equation}\label{solutionsI}
I_1(k^2) = \frac{1}{d-1}P^{\nu\rho}(k)I_{\nu\rho}(k); \quad I_2(k^2) = \frac{1}{k^4(d-1)}(dk^\nu k^\rho - k^2 g^{\nu\rho})I_{\nu\rho}(k)\,.
\end{equation}
Thus, using Eqs.~(\ref{Idecom}) and (\ref{solutionsI}), we obtain from Eq.~(\ref{d1ii}) the following result
\begin{equation}\label{d1survive}
(d_1)_\nu = -\frac{1}{d-1}p_\nu \int_l \frac{(l\cdot k)}{l^2 k^2}[l^2k^2 - (l\cdot k)^2] D(l)D(l+k)\Delta(l)\,,
\end{equation}
where, according to ({\it iii}), we have only written explicitly the contribution that saturates the momentum $p^\rho$ with the metric tensor $g_{\nu\rho}$.

Consider finally the contribution of diagram $(d_2)$. After the shift $l\mapsto -l$, and setting $p=0$ inside the integral, it becomes
\begin{equation}\label{d2i}
(d_2)_\nu = p^\rho (k+p)^\beta \int_l  D(l)\Delta(l)\Delta(l+k)P_\rho^\sigma(l)P_\beta^\alpha(l+k)\Gamma^{[0]}_{\nu\sigma\alpha}\,.
\end{equation}
It is then elementary to show that
\begin{eqnarray}\label{tensorial}
p^\rho (k+p)^\beta P_\rho^\sigma(l)P^\alpha_\beta(l+k) \Gamma^{[0]}_{\nu\sigma\alpha} &=& 2p^\rho \frac{l_\nu l_\rho}{l^2(l+k)^2}[l^2 k^2 -(l\cdot k)^2 + (l+k)^2(l\cdot k)] \nonumber \\
&+& 2p_\nu \frac{[(l\cdot k)^2 - l^2 k^2]}{(l+k)^2} + {\cal O}(p)(k+p)_\nu\,,
\end{eqnarray}
and, therefore, the part of diagram $(d_2)$ to be saved is
\begin{equation}\label{d2ii}
(d_2)_\nu = 2p_\nu \int_l \frac{[(l\cdot k)^2 - l^2 k^2]}{(l+k)^2} D(l)\Delta(l)\Delta(l+k) + 2p^\rho Q_{\nu\rho}(k)\,,
\end{equation}
where we have defined the integral
\begin{equation}\label{Qint}
Q_{\nu\rho}(k) = \int_l \frac{l_\nu l_\rho}{l^2(l+k)^2}[l^2 k^2 - (l\cdot k)^2 + (l+k)^2(l\cdot k)] D(l)\Delta(l)\Delta(l+k)\,.
\end{equation}

One observes at this point 
that the first term in Eq.~(\ref{d2ii}) is already saturated by $p_\nu$ and may be assigned to the form factor $A(k)$ without further considerations. 
On the other hand, decomposing the integral Eq.~(\ref{Qint}) in the second term as
\begin{equation}\label{Qdecom}
Q_{\nu\rho}(k) = Q_1(k^2)g_{\nu\rho} + Q_2(k^2)k_\nu k_\rho\,,
\end{equation}
with
\begin{equation}\label{Qsolutions}
Q_1(k^2) = \frac{1}{d-1}P^{\nu\rho}(k)Q_{\nu\rho}(k); \quad Q_2(k^2) = \frac{1}{k^4(d-1)}(dk^\nu k^\rho - k^2 g^{\nu\rho})Q_{\nu\rho}(k)\,.
\end{equation}
we obtain from Eq.~(\ref{d2ii}) the result 
\begin{equation}\label{d2survive}
(d_2)_\nu = \frac{2}{d-1}p_\nu \int_l \frac{[l^2 k^2 - (l\cdot k)^2]}{l^2k^2(l+k)^2}[(l+k)^2(l\cdot k) - (l\cdot k)^2 - (d-2)l^2k^2] D(l)\Delta(l)\Delta(l+k)\,,
\end{equation}
where, as before, we have omitted terms of the type ${\cal O}(p)(k+p)_\nu$.

Once Eqs.~(\ref{d1survive}) and~(\ref{d2survive}) have been derived, we will use Eq.~(\ref{factorA}) 
for projecting out the form factor $A(k)$, as well as Eqs.~(\ref{Euclidean}) and~(\ref{spherical}), 
in order to pass to Euclidean space, and subsequently cast the answer in spherical coordinates. Thus, we arrive at the final result
\begin{eqnarray}\label{Ap01}
A(y) &=& 1 - \frac{\alpha_s C_{\rm A}}{12\pi^2}\int_0^\infty \!\!\!dt \,\sqrt{yt}\,F(t)\Delta(t) \int_0^\pi \!\!\!d\theta'\sin^4\theta'\cos\theta'\left[\frac{F(u)}{u}\right]  \\
&+&\frac{\alpha_s C_{\rm A}}{6\pi^2} \int_0^\infty \!\!\! dt \,F(t)\Delta(t) \int_0^\pi \!\!\! d\theta' \sin^4\theta' \left[ \frac{\Delta(u)}{u}\right][yt(1+\sin^2\theta')-(y+t)\sqrt{yt}\cos\theta'] \nonumber \,.
\end{eqnarray}
Notice that, in this case, $y=k^2$, $u = (l + k)^2$, and $\theta'$ is the angle between $k$ and $l$.

\section{Numerical Results\label{numsec}}

In this section we will carry out a detailed numerical analysis of the equations obtained in the previous sections.
Specifically, in the first subsection we determine $A(0,-p,p)$ by solving the integral equation \1eq{Aeuclidean}, 
using the lattice data of~\cite{Bogolubsky:2007ud} as input for the gluon propagator $\Delta(q)$ 
and the ghost dressing function $F(q)$ appearing in it. The solution obtained is then compared with the 
lattice data of~\cite{Ilgenfritz:2006he,Sternbeck:2006rd}.
In the second subsection, 
we solve numerically the coupled system formed by the integral equations of the ghost dressing function (\ref{tt2app}) and
of the ghost-gluon vertex in the soft ghost configuration, given by (\ref{Ap01}). 
The unique external ingredient used when solving this system are the  lattice data for the gluon propagator $\Delta(q)$. 
The solution obtained for $F(q)$ compares very favorably with the lattice data of~\cite{Bogolubsky:2007ud}. 

\begin{figure}[!t]
\begin{minipage}[b]{0.45\linewidth}
\centering
\hspace{-1cm}
\includegraphics[scale=0.55]{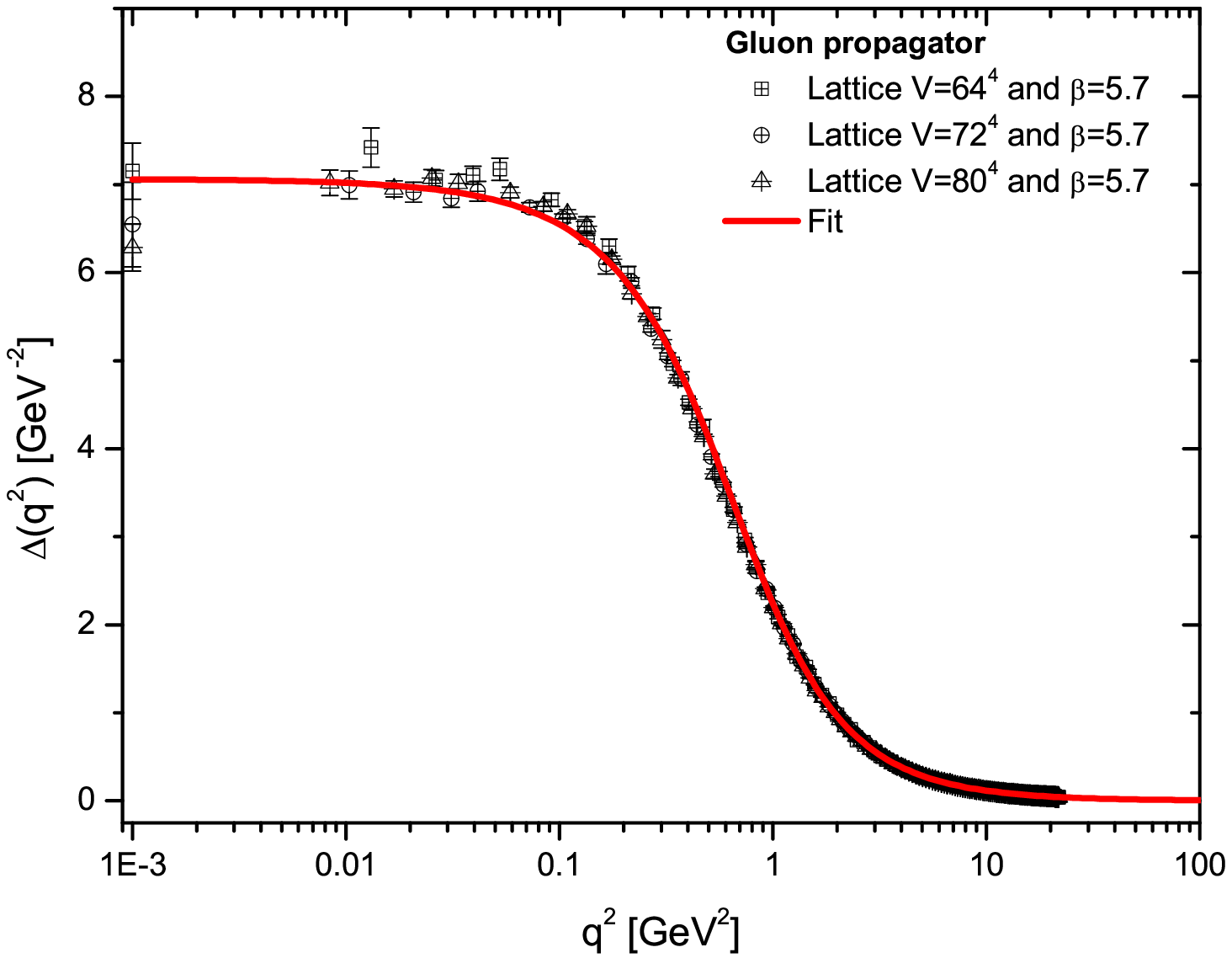}
\end{minipage}
\hspace{0.5cm}
\begin{minipage}[b]{0.50\linewidth}
\hspace{-1.5cm}
\includegraphics[scale=0.55]{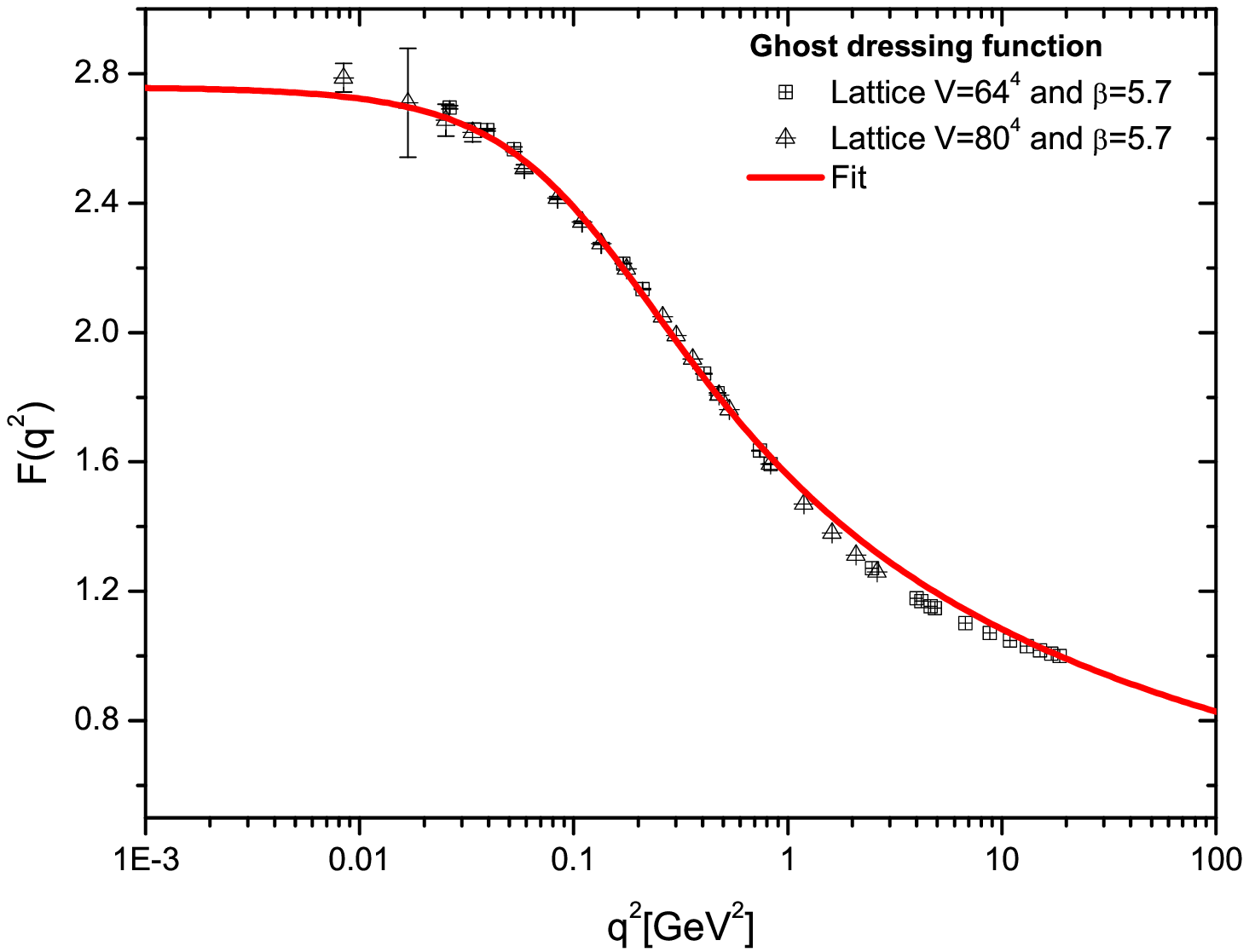}
\end{minipage}
\vspace{-1.0cm}
\caption{Lattice results for the gluon propagator, $\Delta(q)$, (left panel) and ghost dressing, $F(q)$, (right panel)
obtained in  Ref.~\cite{Bogolubsky:2007ud} and renormalized at $\mu=4.3$ GeV. The (red) continuous curves represent the corresponding fits
for the lattice data.}
\label{fig1}
\end{figure}

\subsection{Solution for the soft gluon configuration}

The integral equation~(\ref{Aeuclidean}) is solved through an iterative process, 
using as input for the gluon propagator and the ghost dressing function the 
data obtained from the $SU(3)$ quenched simulations of~\cite{Bogolubsky:2007ud}, shown 
in  Fig.~\ref{fig1}. Note that the lattice data shown have been renormalized at 
$\mu=4.3$ GeV, within the MOM scheme. 
The value of $\alpha_s$ that corresponds to this value of $\mu$ may be 
obtained from the higher-order calculation presented in~\cite{Boucaud:2008gn}; 
specifically, we have that $\alpha_s(\mu)=0.22$.

\begin{center}
\begin{figure}[!ht]
\includegraphics[scale=0.6]{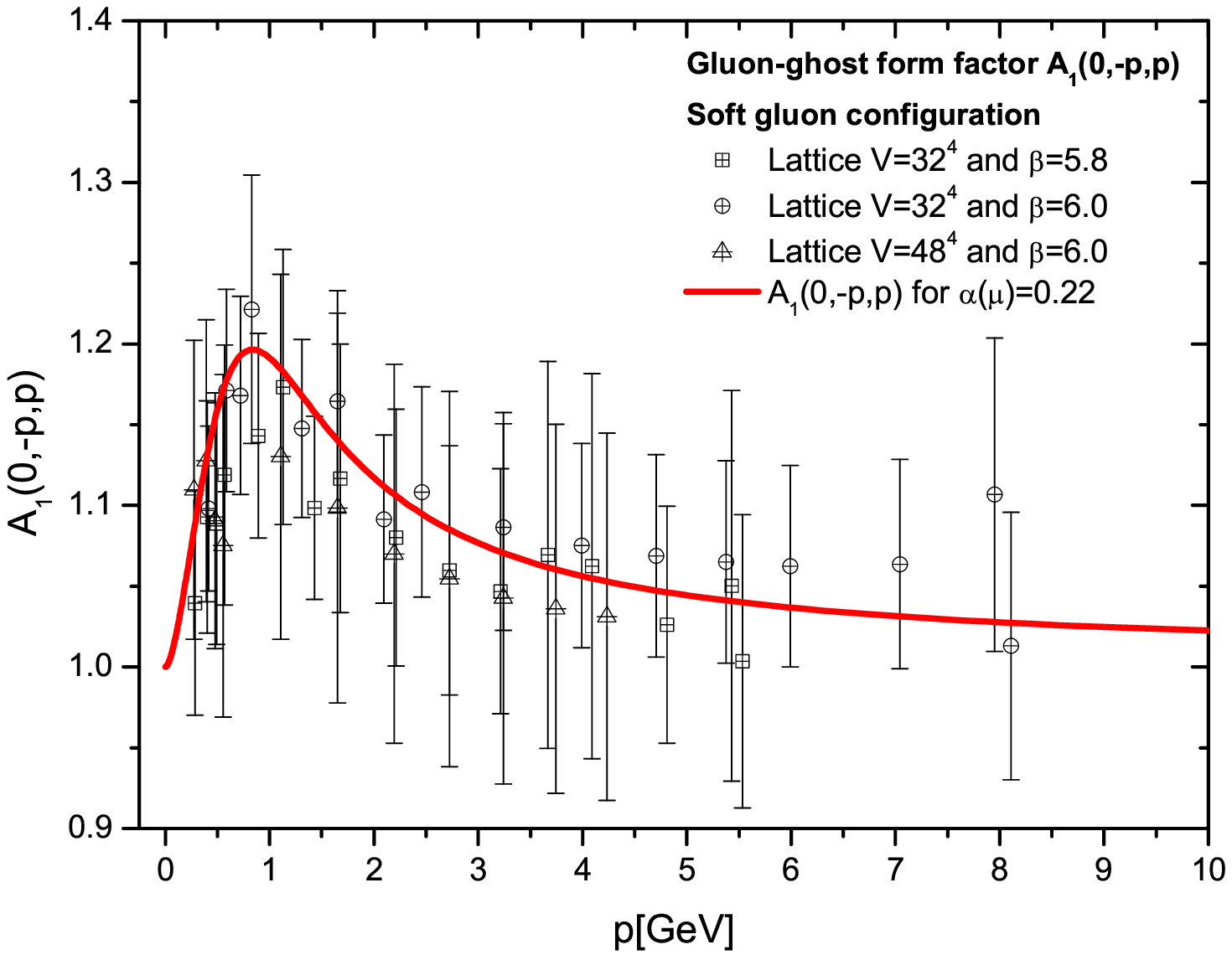}
\caption{Numerical result for $A(0,-p,p)$, obtained from \1eq{Aeuclidean} when $\alpha_s(\mu)=0.22$.} 
\label{fig2}
\end{figure}
\end{center}

The (red) continuous line in Fig.~\ref{fig2} represents the corresponding solution for $A(0,-p,p)$.
We clearly see that $A(0,-p,p)$ 
develops a sizable peak around the momentum region of
\mbox{$830$ MeV}. In addition, as had been anticipated in the subsection~\ref{softh},
we confirm numerically that $A$ indeed  assumes its tree level value
when $p\to 0$, {\it i.e.},  $A = 1$. It is also
interesting to notice that,  in the ultraviolet limit, the form factor gradually
approaches its tree level value. 

In Fig.~\ref{fig2}, we compare our numerical results with the  
corresponding lattice data obtained
in Ref.~\cite{Ilgenfritz:2006he,Sternbeck:2006rd} for this particular kinematic configuration. 
Although, the error bars are rather sizable, we clearly see that 
our solution follows the general structure of
the data. In particular, notice that both 
peaks occur in the same intermediate region of momenta.
Evidently, $A(0,-p,p)$ receives a significant non-perturbative correction, 
deviating considerably from its tree level value.

\subsection{ \label{system} The coupled system: ghost SDE and ghost-gluon vertex.} 

\begin{figure}[!t]
\begin{minipage}[b]{0.45\linewidth}
\centering
\hspace{-1cm}
\includegraphics[scale=0.55]{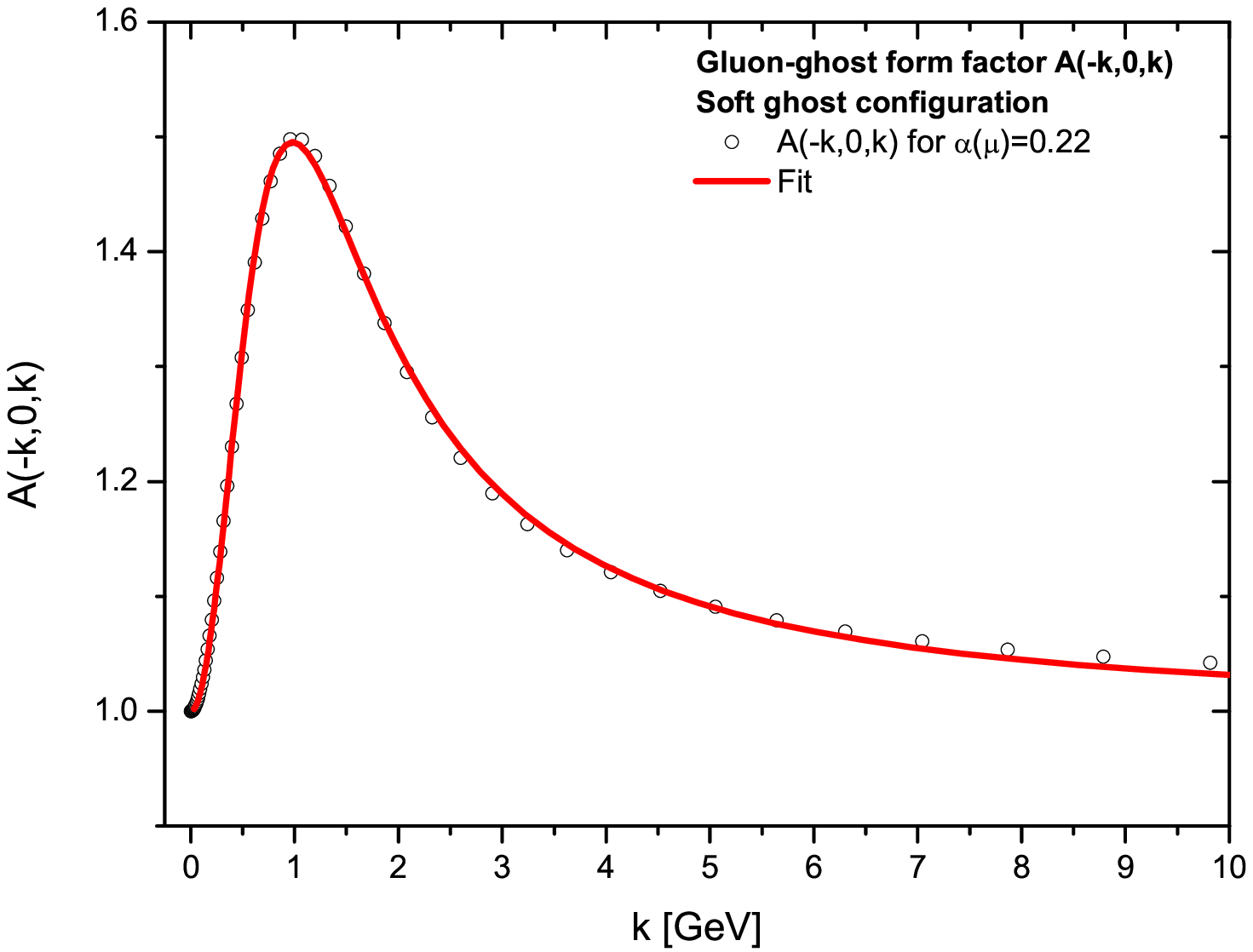}
\end{minipage}
\hspace{0.5cm}
\begin{minipage}[b]{0.50\linewidth}
\hspace{-1.5cm}
\includegraphics[scale=0.55]{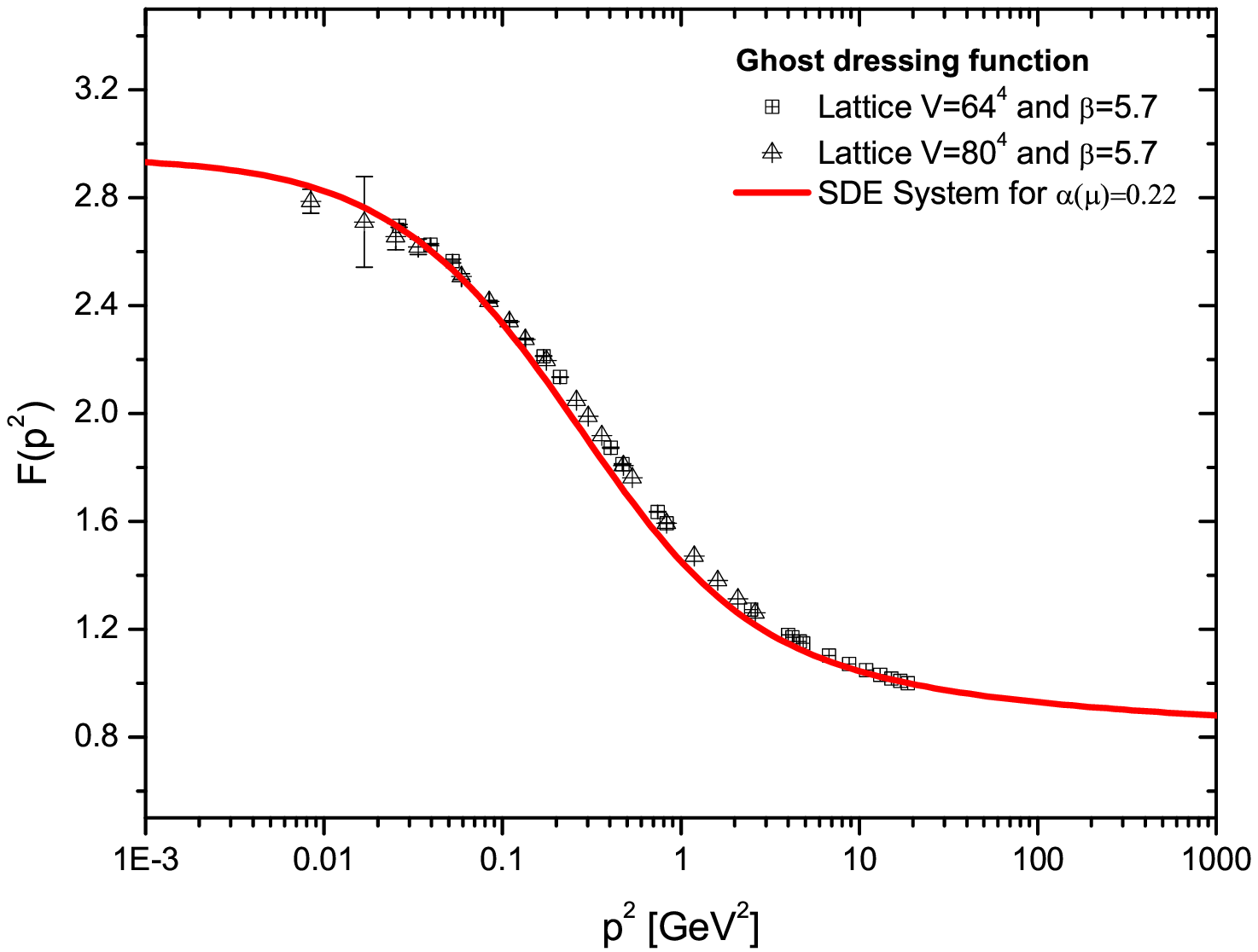}
\end{minipage}
\vspace{-1.0cm}
\caption{ 
{\it Left panel}: The form factor $A(-k,0,k)$ (circles) and
the fit given by \1eq{fit} (red continuous line).  
{\it Right panel}: The numerical solution of $F(p)$ (red continuous line) compared  with 
the lattice data of Ref.~\cite{Bogolubsky:2007ud}. Note that the value of $\alpha_s$ used when solving the system is $\alpha_s(\mu)  =0.22$.}
\label{fig3}
\end{figure}

In this subsection we present the central result of the present article, 
namely the modifications induced to the ghost dressing function 
by the inclusion of a non-trivial structure for the corresponding ghost-gluon vertex.

To that end, after passing to the Euclidean space and introducing spherical coordinates, 
using \1eq{Euclidean} and  \1eq{spherical}, we obtain from  
Eqs.~(\ref{tt2app}) and  (\ref{Ap01}) the expressions 
\begin{eqnarray}
\label{sde_eucl}
F^{-1}(x) = 1 &-& \frac{\alpha_s C_{\rm A}}{2\pi^2}\int_0^\infty \!\!\!dy y \,\Delta(y)A(y) \int_0^\pi \!\!\!d\theta\sin^4\theta\left[\frac{F(z)}{z} - \frac{F(z^{\prime})}{z^{\prime}}\right] \,.
\end{eqnarray}
and
\begin{eqnarray}\label{Ap0}
A(y) &=& 1 - \frac{\alpha_s C_{\rm A}}{12\pi^2}\int_0^\infty \!\!\!dt \,\sqrt{yt}\,F(t)\Delta(t) \int_0^\pi \!\!\!d\theta'\sin^4\theta'\cos\theta'\left[\frac{F(u)}{u}\right]  \\
&+&\frac{\alpha_s C_{\rm A}}{6\pi^2} \int_0^\infty \!\!\! dt \,F(t)\Delta(t) \int_0^\pi \!\!\! d\theta' \sin^4\theta' \left[ \frac{\Delta(u)}{u}\right][yt(1+\sin^2\theta')-(y+t)\sqrt{yt}\cos\theta'] \nonumber \,,
\end{eqnarray}
where now $z = (k+p)^2$, $z^{\prime} = (k + \mu)^2$ and $\mu$ is the renormalization point 
introduced within the MOM scheme, {\it i.e.}, by requiring that $F^{-1}(\mu^2) =1$.

We next solve the above system iteratively, 
using again the lattice data for  $\Delta(q)$ and $\alpha_s(\mu)=0.22$ as input. 
The results for $F(p)$ and  $A(-k,0,k)$ are shown in~Fig.~\ref{fig3}.

On the left panel of~Fig.~\ref{fig3}, the curve in circles represents the result for $A(-k,0,k)$.
Evidently, $A$  develops a peak 
in the intermediate region of momenta, in a way similar to the case discussed in the previous subsection. 
In this case the maximum
of the peak occurs around \mbox{$1$ GeV}, and once more, in the infrared and ultraviolet limits $A(-k,0,k)$
assumes its tree-level value.  

On the same panel we show a fit for $A(-k,0,k)$, represented by the (red) continuous curve, whose functional form is given by
\be
A(-k,0,k)= 1 + \frac{ak^2}{[(k^2+b)^2 + c]\ln\big(d + k^2/k_0^2\big)} \,,
\label{fit}
\ee
with the following values for the fitting parameters \mbox{$a=0.68\,\mbox{GeV}^2$}, \mbox{$b=0.72\,\mbox{GeV}^2$}, \mbox{$c=0.29\,\mbox{GeV}^4$}, \mbox{$d=9.62$} and \mbox{$k_0^2=1\,\mbox{GeV}^2$}.

On the right panel of~Fig.~\ref{fig3},
we compare our numerical result for $F(p)$ (red continuous curve) with the  
corresponding lattice data of Ref.~\cite{Bogolubsky:2007ud}, observing a rather notable agreement. 
We emphasize that, contrary to what happens when the bare vertex is used (see Fig.~\ref{oldsol}), 
the accuracy achieved here does not rely on the artificial enhancement of 
the value of the coupling; the latter, as mentioned above, was kept at its  
standard value predicted from general MOM considerations. 

It is important to realize that,
although $A$  does not provide a sizable support for ghost SDE in the deep infrared, the contribution that 
it furnishes in the region of intermediate momenta is sufficient  for increasing 
the saturation point from $F(0)=1.67$ to $F(0)= 2.95$ (Figs.~\ref{oldsol} and \ref{fig3}, respectively).
This observation suggests that the ghost SDE is particularly sensitive to the values of its ingredients 
at momenta around two to three times the  QCD mass scale.

\section{\label{concl}Conclusions}

In the present work we have considered the ``one-loop dressed'' approximation of the SDE that 
governs the evolution of the ghost-gluon vertex. In particular, we have focused 
on the dynamics of the form factor denoted by $A$,  
which is the one that survives in the SDE for ghost dressing function, in the LG. 
The vertex SDE has been evaluated for two special kinematic configurations, one of them corresponding 
to the well-known Taylor limit. 
When coupled to the SDE of the  ghost, the contribution of this particular 
form factor accounts for the missing strength of the associated kernel, 
allowing one to reproduce the lattice results rather accurately, using the standard value of 
the gauge coupling constant.

The fact that, despite the truncation implemented on the vertex SDE, 
we finally obtained a rather  good agreement with the lattice, 
hints to the 
possibility that  
the omitted terms are numerically subleading, at least in the case of 
the special kinematic configurations considered.
It might be interesting to pursue this point further. Specifically, in the present analysis 
the terms proportional to the second form-factor, denoted by $B$,
have been automatically discarded, precisely because they do not contribute to the 
ghost SDE. However, given that both form factors participate in the fundamental relation of \1eq{Taylteor}, 
one might consider the possibility of keeping these terms throughout the calculation, and then 
checking explicitly to what extent \1eq{Taylteor} is satisfied in the present approximation.

Recently, the study of the 
effects that the dynamical quarks induce on some of the fundamental Green's functions of QCD
has received particular attention, both from the point of view of unquenched lattice simulations~\cite{Ayala:2012pb},
as well as by means of an SDE-based approach~\cite{Aguilar:2012rz}. 
In particular, lattice simulations reveal  
that the inclusion of light active quarks results in a considerable suppression 
in the deep infrared and intermediate momentum region of the gluon propagator.
This characteristic feature has been  
firmly established also within the SDE framework of~\cite{Aguilar:2012rz}. 
On the other hand, the 
unquenched ghost dressing function simulated on the lattice 
suffers minimal changes from the inclusion of 
quarks~\cite{Ayala:2012pb}; this property has also been anticipated within the aforementioned 
SDE analysis~\cite{Aguilar:2012rz}, 
as a direct consequence of the fact that, in the case  of $F$,  
the quark-loops enter as ``higher-order'' effects. 
In addition, it is well-known that the value of the MOM coupling, 
$\alpha(\mu)$, increases in the presence of quark loops.

It would be, therefore, interesting, to study the combination of these competing effects systematically, 
including the vertex equation for $A$, derived here. In particular, 
the nonlinear nature of the corresponding integral equations converts this combined analysis 
into a rather challenging problem.
Specifically, the changes induced to the integral equation for $A$, due to the aforementioned suppression of the 
gluon propagators entering in it, must be compensated, to a considerable level of accuracy,
by the corresponding increase in the coupling constant, 
in order to finally obtain the rather minor change observed in $F$. 
We hope to be able to carry out such a study in the near future.

\acknowledgments 

The research of J.~P. is supported by the Spanish MEYC under 
grant FPA2011-23596. The work of  A.~C.~A  is supported by the 
National Council for Scientific and Technological Development - CNPq
under the grant 306537/2012-5 and project 473260/2012-3,
and by S\~ao Paulo Research Foundation - FAPESP through the project 2012/15643-1.

\end{document}